# The Velocity of Censorship: High-Fidelity Detection of Microblog Post Deletions

Tao Zhu
*zhutao777@gmail.com*
*Independent Researcher*

David Phipps
*Computer Science*
*Bowdoin College*

Adam Pridgen
*Computer Science*
*Rice University*

Jedidiah R. Crandall
*Computer Science*
*University of New Mexico*

Dan S. Wallach
*Computer Science*
*Rice University*

## Abstract

Weibo and other popular Chinese microblogging sites are well known for exercising internal censorship, to comply with Chinese government requirements. This research seeks to quantify the mechanisms of this censorship: how fast and how comprehensively posts are deleted. Our analysis considered 2.38 million posts gathered over roughly two months in 2012, with our attention focused on repeatedly visiting "sensitive" users. This gives us a view of censorship events within minutes of their occurrence, albeit at a cost of our data no longer representing a random sample of the general Weibo population. We also have a larger 470 million post sampling from Weibo's public timeline, taken over a longer time period, that is more representative of a random sample.

We found that deletions happen most heavily in the first hour after a post has been submitted. Focusing on original posts, not reposts/retweets, we observed that nearly 30% of the total deletion events occur within 5–30 minutes. Nearly 90% of the deletions happen within the first 24 hours. Leveraging our data, we also considered a variety of hypotheses about the mechanisms used by Weibo for censorship, such as the extent to which Weibo's censors use retrospective keyword-based censorship, and how repost/retweet popularity interacts with censorship. We also used natural language processing techniques to analyze which topics were more likely to be censored.

## 1 Introduction

Virtually all measurements of Internet censorship are biased in some way, simply because it is not feasible to test every keyword or check every post at small increments of time. In this paper, we describe our method for tracking censorship on Weibo, a popular microblogging platform in China, and the results of our measurements. Our system focuses on a core set of users who are interconnected through their social graph and tend to post about sensitive topics. This biases us towards the content posted by these particular users, but enables us to measure with high fidelity the speed of the censorship and discern interesting patterns in censor behaviors.

Sina Weibo (`weibo.com`, referred to in this paper simply as "Weibo") has the most active user community of any microblog site in China [39]. Weibo provides services which are similar to Twitter, with @usernames, #hashtags, reposting, and URL shortening. In February 2012, Weibo had over 300 million users, and about 100 million messages sent daily [3]. Like Twitter in other countries, Weibo plays an important role in the discourse surrounding current events in China. Both professional reporters and amateurs can provide immediate, first-hand accounts and opinions of events as they unfold. Also like Twitter, Weibo limits posts to 140 characters, but 140 characters in Chinese can convey significantly more information than in English. Weibo also allows embedded photos and videos, as well as comment threads attached to posts.

China employs both backbone-level filtering of IP packets [5, 6, 11, 23, 37, 43] and higher level filtering implemented in the software of, for example, blog platforms [15, 20, 28], chat programs [13, 29] and search engines [30, 41]. Work specific to Weibo [2, 9] is discussed in more detail in Section 2. To our knowledge ours is the first work to focus on how *quickly* microblog posts are removed—on a scale of minutes after they are posted. This fidelity in measurement allows us to not only accurately measure the speed of the censorship, but also to compare censorship speeds with respect to topics, censor methods, censor work schedules, and other illuminating patterns.

What our results illustrate is that Weibo employs "defense-in-depth" in their strategy for filtering content. Internet censorship represents a conflict between the censors, who seek to filter content according to some policy, and the users who are subject to that censorship. Censor-ship can serve to squelch conversations directly as well as to chill future discussion with the threat of state surveillance. Our goal in this paper is to catalog the wide variety of mechanisms that Weibo's censors employ.

This research has several major contributions:

- We describe the implementation of a method that can detect a censorship event within 1–2 minutes of its occurrence. A large amount of Weibo posts are collected constantly via two APIs [26]. There are more than 470 million posts from the public timeline and 2.38 million posts from the user timeline in our database.

- To further understand how the Weibo system can react so quickly in terms of deleting posts with sensitive content, we propose four hypotheses and attempt to support each with our data. We also describe several experiments that shed light on censorship practices on Weibo. The overall picture we illuminate in this paper is that Weibo employs a distributed, defense-in-depth strategy for removing sensitive content.

- Using natural language processing techniques that overcome the usage of neologisms, named entities, and informal language which typifies Chinese social media, we perform a topical analysis of the deleted posts and compare the deletion speeds for different topics. We find that the topics where mass removal happens the fastest are those that are hot topics in Weibo as a whole (*e.g.*, the Beijing rainstorms or a sex scandal). We also find that our sensitive user group has overarching themes throughout all topics that suggest discussion of state power (*e.g.*, Beijing, government, China, and the police).

The rest of this paper is structured as follows. Section 2 gives some basic background information about microblogging and Internet censorship in China. Then Section 3 describes the methods we used for our measurement and analysis, followed by Section 4 that describes the timing of censorship events. Section 5 introduces the natural language processing we applied to the data and presents results from topical analysis. Finally, we conclude with a discussion of various Weibo filtering mechanisms in Section 6.

## 2 Background

Starting from 2010, when microblogs debuted in China, not only have there been many top news stories where the reporting was driven by social media, but social media has also been part of the story itself for a number of prominent events [21, 38], including the protests of Wukan [33], the Deng Yujiao incident [32], the Yao Jiaxin murder case [35], and the Shifang protest [36]. There have also been events where social media has forced the government to address issues directly, such as the Beijing rainstorms in July 2012.

Chinese social media analysis is challenging [27]. One of many concerns that can hinder this work is the general difficulty of mechanically processing Chinese text. Western speakers (and algorithms) expect words to be separated by whitespace or punctuation. In written Chinese, however, there are no such word boundary delimiters. The word segmentation problem in Chinese is exacerbated by the existence of unknown words such as named entities (*e.g.*, people, companies, movies) or neologisms (substituting characters that appear similar to others, or otherwise coining new euphemisms or slang expressions, to defeat keyword-based censorship) [12]. Furthermore, since social media is heavily centered around current events, it may well contain new named entities that will not appear in any static lexicon [8].

Despite these concerns, Weibo censorship has been the subject of previous research. Bamman *et al.* [2] performed a statistical analysis of deleted posts, showing that the presence of some sensitive terms indicated a higher probability of the deletion of a post. Their work also showed some geographic patterns in post deletion, with posts from the provinces of Tibet and Qinghai exhibiting a higher deletion rate than other provinces. WeiboScope [9] also collects deleted posts from Weibo, but their strategy is to follow all users with a high number of followers. This is in contrast to our strategy which is to follow a core set of users who have a high rate of post deletions, some of which have many followers and some of which have few. The deletion events in these works are measured with a resolution of hours or days. Our system is able to detect deletion events at the resolution of minutes.

## 3 Methodology

To have a better understanding of what the Weibo system is targeting for censorship deletions, and how fast they do so, we have developed a system which collects removed posts on targeted users in almost real time.

### 3.1 Identifying the sensitive user group

In Weibo each IP address and Application Programming Interface (API) has a rate limit for access to the service. This forced us to make a number of engineering compromises, notably focusing our attention where we felt we could find those posts most likely to be subject to censorship. We decided to focus on users who we have seen being censored in the past, under the assumption that they will be more likely to be censored in the future. We call this group of users the *sensitive group*.

We started with 25 sensitive users that we discovered manually, leveraging a list from China Digital Times [4] of sensitive keywords which are not allowed to be searched on Weibo's server. To find our initial sample, we searched using out-dated keywords that were later un-banned. For example, 党产共 (Reverse of 共产党, which means "Communist Party") was found to be banned on 4 April 2011, but found to not be banned on 20 October 2011, which means the we were able to obtain some posts containing 党产共 when we searched for this keyword after 20 October 2011. From the search results, we picked 25 users who stood out for posting about sensitive topics.

Next, we needed to broaden our search to a larger group of users. We assumed that anybody who has been reposted more than five times by our sensitive users must be sensitive as well. We followed them for a period of time and manually measured how often their posts were deleted. Any user with more than 5 deleted posts was added to our pool of sensitive users.

After 15 days of this process, our sensitive group included 3,567 users, and within this group we observed more than 4,500 post deletions daily, including about 1,500 "permission denied" deletions. (See Section 3.3 for discussion on different types of deletion events.) Roughly 12% of the total posts from our sensitive users were eventually deleted. Further, we have enough of these posts to be able to run topical analysis algorithms, letting us extract the main subjects that Weibo's censors seemed concerned with on any given day.

We contrast these statistics with WeiboScope [9], developed at the University of Hong Kong in order to track trends on Weibo concurrently with our own study. The core difference between our work and WeiboScope is that they track a large sample: around 300 thousand users who each have more than 1000 followers. Despite this, they report observing no more than 100 "permission denied" deletions per day. WeiboScope's results, therefore, are perhaps more representative of the overall impact of Weibo's censorship as a fraction of total Weibo traffic, while our work has more resolving power to consider the speed and techniques employed by Weibo's censors.

Because we do not have access to WeiboScope's data, we are limited in our ability to make direct comparisons of our datasets. They did briefly support data downloads, and we extracted their "2,500 last permission denied data" on 20 July 2012. This service has since been closed. Our system went live following user timelines on the same date, giving us a single day from which we might compare our data. For 20 July 2012, WeiboScope observed 54 permission-denied posts, while our system observed 1,056.

(Our own system does not yet support public, real-time downloads of our data, which among other issues could make it easier for Weibo to shut it down. An appropriate means of disseminating real-time results or regular summaries is future work for our group.)

While our methodology cannot be considered to yield a representative sample of Weibo users overall, we believe it is representative of how users who discuss sensitive topics will experience Weibo's censorship. We also believe our methodology enables us to measure the topics that Weibo is censoring on any given day.

### 3.2 Crawling

Once we settled on our list of users to follow, we wanted to follow them with sufficient fidelity to see posts as they were made and measure how long they last prior to being deleted. Our target sampling resolution was one minute.

We use two APIs provided by Weibo, allowing us to query individual user timelines as well as the public timeline[1]. Starting in July 2012, we queried each of our 3,500 users, once per minute, for which Weibo returns the most recent 50 posts. Deleted posts outside of this 50-post window are not detected by our system, meaning that we may be underestimating the number of older posts that get deleted.

We also queried the public timeline roughly once every four seconds, for which Weibo returns 200 recent posts. Half of these posts appear to be 1–5 minutes older than real-time, and the other half are hours older.

Weibo does not support anonymous queries to its servers, requiring us to create fake accounts on the service. Weibo further enforces rate limits both on these users' queries as well as on source IP addresses, regardless of what user account is being used for the query. To overcome these concerns, we used roughly 300 concurrent Tor circuits [24], driven from our research computing cluster. Our resulting data was stored and processed on a four-node cluster using Hadoop and HBase [1].

If and when Weibo might make a concerted effort to block us, it is easy to imagine a ongoing game where they invent new detection strategies and we invent new workarounds. So far, this has not been an issue.

### 3.3 Detecting deletions

An absent post may have been censored, or it may have been deleted for any of a variety of other reasons. User accounts can also be closed, possibly for censorship purposes. Users cannot delete their own account, only the system can delete accounts. We conducted a variety of short empirical tests to see if we could distinguish the different cases. We concluded that we can detect two kinds of deletions.

---

[1]The user timeline returns both original posts and retweeted posts by that user, while the public timeline only returns original posts. Also, the public timeline appears to be only a sampling of the total public traffic.

If a user deletes his or her own post, a query for that post's unique identifier will return a "post does not exist" error. We have observed this same error code returned from censorship events and we refer to these, in the remainder of the paper as *general deletion*. However, there is another error code, "permission denied," which seems to indicate that the relevant database record still exists but has been flagged by some censorship event. We refer to these as *permission-denied deletions* or *system deletions*. In either case, the post is no longer visible to Weibo users.

The ratio of system deletions to general deletions in our user timeline data set is roughly 1:2. In this paper, we generally focus on posts that have been system deleted, because there appears to be no way for a user to induce this state. It can only be the result of a censorship event (*i.e.*, there are no censorship false positives in our system deletion dataset). Because we followed a core set of users who post on sensitive subjects, we did not find it necessary to account for spam in our user timeline dataset.

Our crawler, which repeatedly fetches each sensitive user's personal timeline, is searching for posts that appear and then are subsequently deleted. If a post is in our database but is not returned from Weibo, then we issue a secondary query for that post's unique ID to determine what error message is returned. Ultimately, with the speed of our crawler, we can detect a censorship event within 1–2 minutes of its occurrence.

For each returned post from Weibo, there is a field which records the creation time of the post. The lifetime of a post is the time difference between the time our system detected the post being deleted and the creation time. Therefore a post's lifetime recorded by our system is never shorter than its real lifetime, and never longer than its real lifetime by more than two minutes.

## 4 Timing of censorship

For easier explanation we first give some definitions. A post can be a repost of another post, and can have embedded images. Also other users can repost reposts. If post *A* is a repost of post *B*, we call post *A* a *child post* and post *B* a *parent post*. If post *A* is not a repost of another post, we call post *A* a *regular post*.

Using our user tracking method, from 20 July 2012 to 8 September 2012, we have collected 2.38 million user timeline posts, with a 12.8% total deletion rate (4.5% for system deletions and 8.3% for general deletions). Note that this deletion rate is specific to our users and not representative of Weibo as a whole. With a brief analysis, we found that 82% of the total deletions are child posts, and 75% of the total deletions have pictures either in themselves or in their parent post.

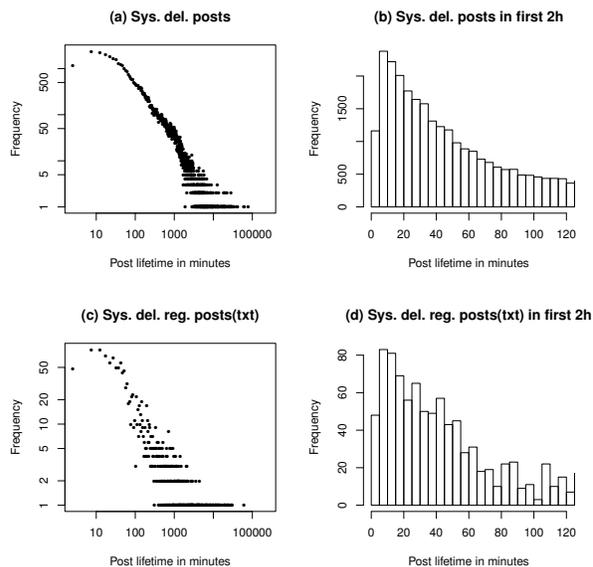

Figure 1: **Lifetime histograms. (a) and (b) are the lifetime histograms of all system deletions. (c) and (d) are the lifetime histograms of regular text-only posts. (a) and (c) show the histogram of the whole lifetime, (b) and (d) only show the first two hours of the lifetime histogram.**

To demonstrate how long a post survives before it gets deleted, we analyze the system deletion data set (see Section 3.3). Figure 1 gives us a big picture of how fast the Weibo system works for censorship purposes. The *x* axes are the length of the lifetime divided into 5-minute bins, and the *y* axes are the count of the deleted posts having the lifetime in the corresponding bin. We note that these figures have the distinctive shape of a power law or long tailed distribution, implying that there is no particular time bound on Weibo's censorship activity, despite the bulk of it happening quickly, and that metrics like mean and median are not as meaningful as they are in a normal distribution.

We can see that the post bins with small lifetimes are large. We zoom into the first 2 hours of data, which is plotted in Figure 1 (c) and (d). This tells us that system deletions start within 5 minutes, the same as text-only regular posts. For both of them, the modal deletion age appears to be between 5–10 minutes.

In our data set, 5% of the deletions happened in the first 8 minutes, and within 30 minutes, almost 30% of the deletions were finished. More than 90% of deletions happened within one day after a post was submitted. This demonstrates why a measurement fidelity on the order of minutes, rather than days, is critical.

Considering the big data set that Weibo has to process, the speed, especially the 5 to 10 minutes peak, is fast, especially considering that the data cannot be processed in a fully automated way. How can the Weibo system find sensitive posts and remove them so quickly? On the other hand, the long tails suggest that sensitive posts can still be deleted even after an extended period. How are those sensitive posts located by the moderators after a month in their huge database? What factors affect a post's lifetime?

In this section, to find the answers to these questions, we propose four hypotheses and then test them against our data. Hypotheses 1 and 2 try to explain how the speed of censorship on Weibo can be so fast. Hypothesis 3 explains why we see the long tails of the post lifetime for censored posts in Figure 1. Hypothesis 4 tells us that the deletion speed does not appear to be strongly related to particular conversation topics, but rather to popular topics (*i.e.*, those that are being discussed on Weibo as a whole according to our public timeline) where our core sensitive users are putting a spin on the discussion that involves themes of government power (see Section 5).

## 4.1 Post lifetime regression

Before we give our hypotheses, we first consider what factors affect a post's lifetime, regardless of the content of the post.

For each post, besides the basic information about the post itself, we also see an embedded picture, if present, as well as a parent post identifier, if it is a repost. Also, we know the number of followers and friends of each user, as well as of any parent post's user.

From the graphs in Figure 1, we decided to experimentally fit a negative binomial regression to it to see which factors affect the lifetime of a post. Table 1 and Table 2 show the results for the regular posts and child posts, respectively. Three asterisks ('***') indicates statistical significance, one asterisk ('*') indicates a coefficient that is not statistically significant, and no coefficient is indicated with a dash ('-'). We can regress the log lifetime for a regular post or a child post via:

$$\ln(\widehat{Regular_{lifetime}}) = Intercept + b_1(P_{HasPic}) \\ + b_2(Friends \#) + b_3(Posts \#)$$

$$\ln(\widehat{Child_{lifetime}}) = Intercept + b_1(P_{HasPic}) \\ + b_2(P.Friends \#) + b_3(P.Posts \#)$$

We examine the effect on post lifetime of: the existence of a picture, the number of friends and followers, and the number of posts sent by this user. We found that, for both regular and child posts, the existence of a picture affects the post's lifetime the most. That is, posts with pictures have shorter lifetimes than posts without pictures. Some of the user attributes, such as number of friends or number of posts, also affect post lifetime. We note that the coefficients for these are relatively small, but for users with large numbers of friends or who write large numbers of posts, these factors can have a significant impact on the speed of that users' posts being censored. However other attributes of a user, such as whether a Weibo user is "verified" by Weibo (*i.e.*, Weibo knows who they are as part of newer Chinese requirements that crack down on pseudonyms unconnected to real world identities) or the number of followers of a user, are not statistically significant factors in a post's lifetime.

Table 1: **Factors affecting post lifetime (regular posts).**

| Factors | Coef | Stat. Sig. |
| --- | --- | --- |
| (Intercept) | 7.41 | *** |
| Has picture | $-4.07 \times 10^{-1}$ | *** |
| Number of friends | $-2.42 \times 10^{-4}$ | *** |
| Number of posts | $-5.23 \times 10^{-5}$ | *** |
| User verified | – | - |
| Number of followers | – | - |

Table 2: **Factors affecting post lifetime (child posts)**

| Factors | Coef | Stat. Sig. |
| --- | --- | --- |
| (Intercept) | 6.27 | *** |
| Parent has picture | $-1.01 \times 10^{-1}$ | *** |
| Parent friends number | $-4.76 \times 10^{-5}$ | *** |
| Parent posts number | $6.84 \times 10^{-6}$ | *** |
| Parent user verified | $2.01 \times 10^{-1}$ | * |
| Parent followers number | – | - |

## 4.2 Hypotheses

As a distributed system with 70,000 posts per minute, Weibo has above a 10% rate of deletion in the public timeline (first observed by Bamman *et al.* [2]; we have seen similar behavior). This high deletion rate can be the result of many processes, including anti-spam features, user deletions, as well as anti-censorship features. Within the deletions that we believe are censorship events, we note that 40% of the deletions in our user timeline data set occur within the first hour after a post has appeared. Clearly, Weibo exerts significant controls over its content.

Before censors deal with the sensitive posts which are already in the system, are there filters which do not allow certain posts to enter the Weibo system? This question leads to our first hypothesis.

**Hypothesis 1** *Weibo has filtering mechanisms as a proactive, automated defense.*

To find out if there are filtering mechanisms, we attempted to post posts containing sensitive words from the China Digital Times [4] and Tao *et al.* [41]. Here we summarize the filtering mechanisms Weibo was found to apply based on our observations.

- **Explicit filtering:** Weibo will inform a poster that their post cannot be released because of sensitive content.

  For example, on 1 August 2012, we tried to post "政法委书记" (Secretary of the Political and Legislative Committee). When we submitted a post with this character string in it, a warning message says "Sorry, since this content violates 'Sina Weibo regulation rules' or a related regulation or policy, this operation cannot be processed. If you need help, please contact customer service."

- **Implicit filtering:** Weibo sometimes suspends posts until they can be manually checked, telling the user that the delay is due to "server data synchronization."

  For example when we submitted the post 'youshenme**falun**debanfa' on the same day, 1 August 2012, Weibo responded with the message "Your post has been submitted successfully. Currently, there is a delay caused by server data synchronization. Please wait for 1 to 2 minutes. Thank you very much." This delay, which frequently takes much longer than the 1–2 minutes suggested by Weibo, was triggered by our use of the substring "falun", pertaining to the Falun Gong religion. In this example, it took more than 5 hours for the post to appear.

- **Camouflaged posts:** Weibo also sometimes makes it appear to a user that their post was successfully posted, but other users are not able to see the post. The poster receives no warning message in this case.

  On 1 August 2012 we submitted a post containing the substring "cgc" (Chen Guangcheng [31]), and received no warning messages, so it seemed to be published successfully to our user. When we tried to access that post from another user account, however, we were redirected to Weibo's error page which claimed the post does not exist.

We found these phenomena to be repeatable. Over the course of our experiments, we selected a number of different subsets of the keyword list published by the China Digital Times [4], trying to post them to Weibo manually. We consistently found all of these same phenomena, although the specific keywords on any list vary over time.

Figure 1 shows that the deletions happen most heavily for a regular post within 5 to 10 minutes of it being posted. While we believe this process to happen largely via automation, it is instructive to estimate how much unaided human labor would otherwise be necessary. Suppose an efficient worker could read 50 posts per minute, including the reposts and figures included in the posts. Then to read Weibo's full 70,000 new posts [34] in one minute, 1,400 simultaneous workers would be needed. Assuming 8 hour shifts, 4,200 workers would then be required. We can imagine that such a staff would have a high error rate, owing to the repetitive nature of their work. Such a labor force would also be relatively expensive compared to automation. *We instead conclude that Weibo must be using a large amount of automation*, perhaps keyword-based as has been found in other systems in China such as TOM-Skype [16]. This is likely complemented with human efforts to evolve and refine the filtering process.

Some of this refinement certainly results from a centralized list of topics. Other refinement may occur internally, through a smaller number of censors who look for users finding new ways to misspell words or otherwise work around existing filters. Our subsequent hypotheses consider how this refinement occurs and delve into how Weibo's automation operates.

**Hypothesis 2** *Weibo targets specific users, such as those who frequently post sensitive content.*

Another way to achieve prompt response to sensitive posts is to track users who are likely to post sensitive content, using techniques similar to what we are doing. The posts from those sensitive users could then be read by moderators more often and more promptly than the posts of other users.

To test this hypothesis, we plotted Figure 2. We grouped users together who have the same number of censorship events occurring to their posts. The *x*-axis is the number of such deletions for each cohort of users. The *y*-axis shows how long these to-be-censored posts live. The clear downward trend is evidence that users with larger deletion frequencies tend to observe faster censorship of their work, supporting our hypothesis.

Even though this figure shows us that the more deletion posts a user has, the faster the users' posts tend to be deleted, we cannot rule out other features which those users have in common and that those features may lead to the fast deletions. For example, they may tend to use the same keywords, post from the same geographical area, use the same kind of client platform, and so on. There is a clear correlation between post lifetime and post deletion counts, but correlation does not imply causation.

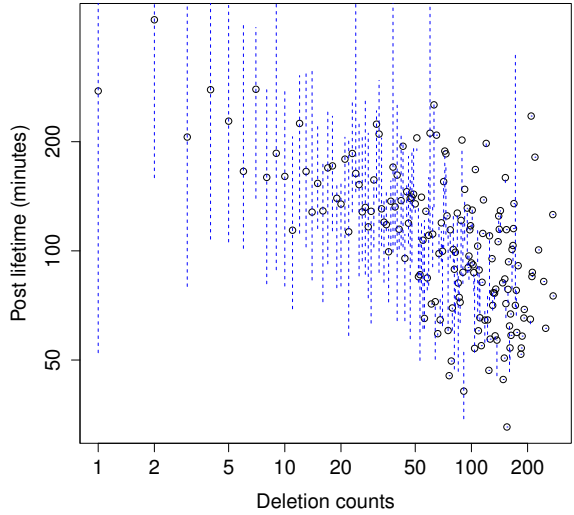
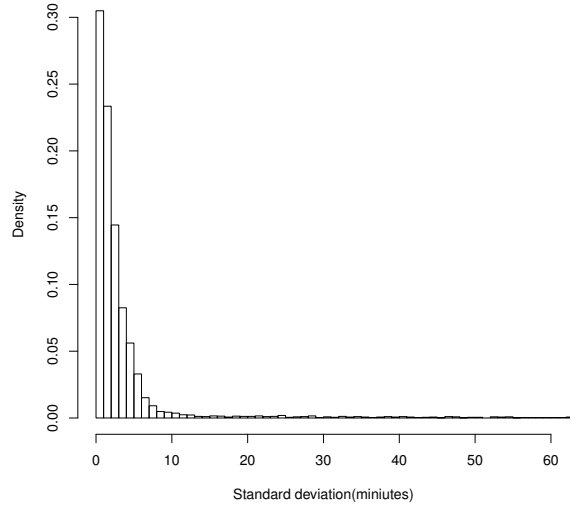

Figure 2: **Users' median post lifetime in minutes *vs.* the number of deletions for that user on a log-log scale. Black circles show the *median* lifetime of posts in the cohort, and the dotted blue bars show the 25%–75% range.**

Figure 3: **Reposts standard deviation histogram.**

If the surveillance keyword list and targeting of specific users were the only mechanisms for removing sensitive posts, then the histograms in Figure 1 would stop at a certain time, say 1 or 2 days. However, 10% of the deletions happen after one day, with some deletions occurring one month or more after the post was posted. Clearly, other mechanisms are in use for these long-tail censorship events, which leads to our next hypothesis.

**Hypothesis 3** *When a sensitive post is found, a moderator will use automated searching tools to find all of its related reposts (parent, child, etc.), and delete them all at once.*

If this hypothesis is true, then the child posts which repost a censored parent post should all be removed at the same time. To test this hypothesis, we plot the histogram of the standard deviation of the deletion time of the posts sharing the same Repost Identification Number (rpid) in Figure 3. In our system deleted posts dataset, over 82% of reposted posts have a deletion time standard deviation of less than 5 minutes, meaning that a sensitive post is detected and then most of the other posts in a chain of reposts are immediately deleted.

There are outliers with standard deviations as high as 5 days which suggest that the mass deletion strategy mentioned here is not the only method Weibo employs to delete sensitive reposts. This leads to our next hypothesis.

**Hypothesis 4** *Deletion speed is related to the topic. That is, particular topics are targeted for deletion based on how sensitive they are.*

We performed topical analysis on the deleted posts. The topical analysis methods we use are described in Section 5.1. Here, to save space, we only list the top topic in Table 3. (For further topical discussion, please refer to our technical report [42].) The third column is the response time for the censor to discover a sensitive topic. Specifically, the response time here refers to the period between the time when the first post on this topic appeared in our user timeline data set and the time when the Weibo system starts to delete the posts on this topic heavily. These times were identified through manual analysis. Even when a topic is still being actively censored, it does not necessarily disappear. People may still discuss the topic only to have their posts deleted. That is why some topics appear twice or more in the table. When a topic showed up again, there is no response time for it and we indicate this with a dash ('-').

The main five topics extracted by Independent Component Analysis (ICA, see Section 5) are: Qidong, Qian Yunhui, Beijing Rainstorm, Diaoyu Island[2] and Group Sex. From Table 3, we can see that these topics have a relatively short lifetime compared to other topics. These five topics were also hot topics in our public timeline during this period.

---

[2]Diaoyu Island is the number 3 top topic on 16 August.

Table 3: **Blocked topics.**

| Date | Top 1 | Response Time (hours) |
|---|---|---|
| 7-20 | Support Syrian rebels | 21.32 |
| 7-21 | Lying of gov. (Jixian) | 12.20 |
| 7-22 | Beijing rainstorms | 2.55 |
| 7-23 | Beijing rainstorms (Subway) | 1.62 |
| 7-24 | Beijing rainstorms[a] | 2.65 |
| 7-25 | Beijing rainstorms (Fangshan) | 2.58 |
| 7-27 | Beijing rainstorms (37 death) | 0.82 |
| 7-28 | Qidong | 1.18 |
| 7-29 | Qidong (Japanese reporter) | 2.25 |
| 7-30 | Complain gov. (Zhou Jun) | 5.73 |
| 7-31 | Judicial independence | 2.00 |
| 8-01 | Complain gov. (Hongkong) | 45.30 |
| 8-02 | Freedom of speech | 7.35 |
| 8-03 | Qidong (Block the village) | 31.58 |
| 8-04 | One-Child Policy Abuse | 33.42 |
| 8-05 | Human Rights News | 24.63 |
| 8-07 | Qian Yunhui Accident | 10.87 |
| 8-08 | Qian Yunhui Accident | – |
| 8-09 | Group sex | 0.78 |
| 8-10 | RTL[b] | 3.65 |
| 8-11 | Tang Hui | 33.42 |
| 8-12 | Group sex | – |
| 8-13 | Corpse Plants in Dalian | 532.50 |
| 8-14 | Hongkong | 70.98 |
| 8-15 | Corpse Plants in Dalian | – |
| 8-16 | Corpse Plants in Dalian | – |
| 8-17 | Complain gov. (North Korea) | 19.83 |
| 8-18 | Zhou Kehua (faked) | 16.37 |

[a] Refuse to donate for Beijing rainstorms.
[b] Re-education through labor.

This suggests that when sensitive users and a large number of regular Weibo users are discussing the same general topic, *i.e.*, the topic is popular in both the user timeline and public timeline, then extra resources are devoted to finding and deleting such posts[3]. In Section 5 we will show that the sensitive users in the user timeline combine topics with common themes related to state power (Beijing, government, China, country, police, and people). This suggests that the censors consider the combination of these themes with generally popular topics to warrant extra resources.

---

[3] We have not ruled out other possibilities in our study, however, such as that such topics are viewed by many users and therefore more likely to be reported by regular users.

## 5 Topic extraction

Even though we are following a relatively modest number of Weibo authors, the volume of text we are capturing is still too much to process manually. We need automatic methods to classify the posts that we see, particularly those which are deleted.

Automatic topic extraction is the process of identifying important terms in the text that are representative of the corpus as a whole. Topic extraction was originally proposed by Luhn [19] in 1958. The basic idea is to assign weights to terms and sentences based on their frequency and some other statistical information.

However, when it comes to microblog text, standard language processing tools become inapplicable [18, 40]. Microblogs typically contain short sentences and casual language [7]. Unknown words, such as named entities and neologisms often cause problems with these term-based models. It can be especially challenging to extract topics from Asian languages such as Chinese, Korean, and Japanese, which have no spaces between words.

We applied the Pointillism approach [27] and TF*IDF to extract hot topics. In the Pointillism model, a corpus is divided into n-grams; words and phrases are reconstructed from grams using external information (specifically, temporal correlations in the appearance of grams), giving the context necessary to manage informal uses of the language such as neologisms. Salton's TF*IDF [10] assigns weights to the terms of a document based on the terms' relative importance to that document compared to the entire corpus.

We next explain how these techniques work together.

### 5.1 Algorithm

TF*IDF is a common method to determine the importance of words to a document in a corpus. The TF*IDF value in our case is calculated as:

$$f(t, d_{day}) \times log \frac{\textit{Total number of posts for the month}}{f(t, d_{month})}$$

Here, $f(t,d)$ means the frequency of the term $t$ in document $d$. We use trigrams as $t$, and documents $d$ are sets of posts over a certain period of time. $d_{day}$ is the deleted posts we caught on day *day*. We use the posts of July, 2012 in the public timeline as IDF. $f(t, d_{month})$ is the frequency of term $t$ in the public timeline in July, 2012.

First we calculate TF*IDF scores for all trigrams that have more than 20 occurrences in a day. The top 1000 trigrams with the highest TF*IDF score will be fed to our trigram connection algorithm, hereafter "Connector." We call these top 1000 trigrams the *1000-TFIDF list*.

To connect trigrams back into longer phrases, Connector finds two trigrams which have two overlapping characters. For instance, if there are ABC and BCD, Connector will connect them to become ABCD. Sometimes there is more than one choice for connecting trigrams, *e.g.*, there could also be BCE and BCF. Sometimes trigrams can even form a loop. To solve these problems, we first build directed graphs for the trigrams with a high TF*IDF score. Each node is a trigram, and edges indicate the overlap information between two trigrams. For example, if ABC and BCD can be connected to make ABCD, then there is an edge from 'ABC' to 'BCD'. After all trigrams are selected, we use DFT (Depth First Traversal) to output the nodes. During the DFT we check to see if a node has been traversed already. If so we do not traverse it again. After the graphs have been traversed, we obtain a set of phrases.

For example, the Connector output of the third most popular topic on 4 August 2012 is:

> 1.头骨进京鸣冤。河北广平县上坡村76岁的农民冯虎，其子在19
> skull go Beijing to redress an injustice. The son of a 76 year old farmer Fenghu, from Shangpo village, Guangping city, Hebei province, was ... at 19
> 2.头骨进京鸣冤。冯出示的头骨赴京鸣...
> skull go Beijing to redress an injustice. The skull shown by Feng go Beijing to redress an injustice...
> 3.头骨进京鸣冤。冯出示的头骨前额有一大窟窿，他...
> skull go Beijing to redress an injustice. There is a big hole on the skull shown by Feng, he...
> 4.头骨进京鸣冤。冯出示的头骨前额有一个无罪的公民...
> skull go Beijing to redress an injustice. There is a innocent citizen on the skull shown by Feng, he...
> 5.头骨进京鸣冤。冯出示的头骨进...
> skull go Beijing to redress an injustice. The skull shown by Feng enter...
> 6.头骨进京鸣冤。冯出示的头等舱
> skull go Beijing to redress an injustice. The first class seat shown by Feng...
> 7.【華聯社電】上访15年 老父携儿头骨...
> Chinese Community report: petition 15 years, old father bring the skull of his son...

In this example, the 7 outputs of Connector are translated in English, which is written in the next line after the original Chinese phrase. Outputs 4 and 6 are incorrectly connected. This is because the same trigrams are shared by different stories that have high TF*IDF scores on the same day. This problem can be solved by examining the cosine similarity of the frequency of occurrence of the first and the last trigram for each result.

Cosine similarity is used to judge whether two trigrams have correlated trends.

$$cos.Sim = \frac{<A_i, B_i>}{\sqrt{\sum_{i=1}^{n} A_i^2} \times \sqrt{\sum_{i=1}^{n} B_i^2}}$$

where $<,>$ denotes an inner product between two vectors. For details, please refer to Song *et al.* [27].

From the connected sentences, listed above, we can begin to understand the general events that are driving major sensitive topics of discussion on Weibo. Table 3 lists the top topics of the deleted posts from 20 July 2012 to 18 August 2012. (A computer failure prevented us from collecting data on 6 August 2012.) Note that we just translated the posts from each topical cluster, we have not confirmed the veracity of any of the claims of the Weibo users' posts that we translated.

Interestingly, besides named entities, we also extracted three neologisms. They are 李W阳 (Li Wangyang, from 李旺阳), 六圖四 (June Fourth, from 六四), 胡()涛 (Hu Jintao, from 胡锦涛, replacing the middle character with open- and close-parentheses), and 启-东, 启\东 and 启/东 (Qidong, from 启东, inserting punctuation between the two characters). These neologisms became popular enough that they stood out in our TF*IDF analysis.

### 5.2 Hot sensitive topics

Table 3 tells us the top topic for each day in terms of having the highest TF*IDF scores—however, it does not tell us which topics among these have been discussed for the longest period of time by our users. Also, are there some common themes behind those separate topics?

Here are the top 50 words which have appeared in the *1000-TFIDF list* most frequently from 20 July 2012 to 20 August 2013, manually translated to English:

Beijing City, Liu Futang, secretary, Lujiang County, Guo Jinlong, Qian Yunhui, City Government, Zhou Kehua, Red Cross, Diaoyu Island, subprefect, water drain, ordinary people, taxpayer, Fangshan district, Hagens, local police station, office, Beijing, Qidong, government, China, Japan, citizen, county's head commissioner, reporter, mayor, corrupt official, freedom, country, restrain, keyhole report, wrist watch, police, national, recommend, American, repression, patriotic, democratic, corpses, people, donation, cancel, opinion, reeducation through labor, abolition, truck[4]

We used Independent Component Analysis (ICA) to extract "independent signals" from those most important terms shown above. ICA [14] is a method to separate a linearly mixed signal, *x*, into mutually independent components, *s*.

---
[4]For clarity, we have elided close variants on *China*, *Japan*, and *Beijing* from this list.

Let $X = [x_1, x_2, ..., x_m]^T$ be the observation mixture matrix, consisting of *m* observed signals $x_i$. Since X is the linear composition of the independent components, *s*, X can be modeled as:

$$X = AS = \sum_{i=1}^{m} a_i s_i$$

*A*, the mixing matrix, gives the coefficients for linear combinations of the independent signals, the rows of *S*.

Here, each word is represented by a row vector of length 864 (36 × 24), which contains the 36 days worth of hourly frequency from 22 July 2012 to 2 September 2012. The 50 × 864 matrix *X* is fed to an ICA program [25]. The number of independent components number is set to 5, which retains almost 100% of the eigenvalues.

There are six words that appear in almost every independent signal: Beijing, government, China, country, policeman, and people. This means that the sensitive user group in our user timeline has these general themes that cut across the many individual topics that they discuss, which may explain why their posts are often subject to censorship.

## 6 Discussion

Weibo appears to have a variety of other mechanisms that do not fit neatly into our hypotheses, but which are interesting to discuss. We first consider other aspects of Weibo's filtering, then we look at diurnal (time-of-day) censorship behaviors, and finally we try to synthesize some of our observations.

### 6.1 Weibo's filtering mechanisms

Sina Weibo has a complex variety of censorship mechanisms, including both proactive and retroactive mechanisms. Here we summarize the mechanisms Weibo may apply. Proactive mechanisms, as we discussed in Hypothesis 1, may include: explicit filtering, implicit filtering, and camouflaged posts. Retroactive mechanisms for removing content that has already been released may include:

- **Backwards reposts search:** In our deleted posts dataset over 82% of reposted posts have a standard deviation of less than 5 minutes for deletion time, meaning that a sensitive post is detected and then most of the other posts in a chain of reposts are then deleted (Hypothesis 3).

- **Backwards keyword search:** We also observed that Weibo sometimes removes posts retroactively in a way that causes spikes in the deletion rate of a particular keyword within a short amount of time.

  Here, we give two examples (兲朝 and 37人), out of many that we witnessed, with a strong spike in the deletion of posts containing that keyword.

  We first consider 兲朝, Tian Chao, a neologism for "Celestial Empire" where 兲 is an alternate form for 天; the substitute character is visually similar to the original and also appears to be constructed from the two distinct characters 王八, meaning "bastard."). The frequency of 兲朝 in deleted posts, day by day, is the sequence (6,3,0,0,2,2,0,3,0,2,3,3,2,1,2,0,0,1,0,0,0,5,4,4,2,**14**,3,6,4) respectively from 28 July 2012 to 25 August 2012. There is a concentrated deletion (14 censorship events) of posts with this word within several minutes on 22 August 2012, impacting posts that were several weeks old at the time. It is likely that a censor discovered this new phrase and ordered it globally expunged.

  Another example is the keyword 37人 (37 people). There are 44 posts containing this keyword, which were created from 2 days to 5 days before the censorship event, all removed together within 5 minutes (03:25 to 03:30 27 July 2012). Those 44 posts are from different users, have no common parent posts, and have no common pictures. The only plausible explanation for this concentrated deletion would appear to be a keyword-based deletion. The deletion time at 3:25am Beijing time also strongly suggests that there are moderators working in the early morning. To understand this workforce and its distributed nature, we perform further analysis in Section 6.2.

- **Monitoring specific users:** Hypothesis 2 shows a clear preference for Weibo's censors to pay more attention to users who seemingly like to discuss censored topics.

- **Account closures:** Weibo also closes users' accounts. There were over 300 user accounts closed by the system from our sensitive user group (out of over 3,500 users) during the roughly two month period while when we collected data for their user timelines.

- **Search filtering:** To prevent users from finding sensitive information on weibo.com, Weibo also has a frequently updated list of words [4] which cannot be searched.

- **Public timeline filtering:** We believe that sensitive topics are filtered out of the public timeline. This

filtering appears to be limited to only general topics that have been known to be sensitive for a relatively long time. In this paper all major results are based on the user timeline, we only use the public timeline for general results about major trending topics in Weibo.

- **User credit points:** In May 2012, Sina Weibo announced a "user credit" points system [22] through which users can report sensitive or rumor-based posts to the administrators. We do not know the extent to which the point system interacts with the censorship mechanisms that we have already described. It is possible that these reports "bubble up" and help Weibo tune its automated filters, but we have no way to observe this.

## 6.2 Time-of-day behavior

In our data, the time at which the censors are working and deleting posts correlates more with the usage patterns of regular users than with a typical day-time work schedule (*e.g.*, 8am to 5pm Beijing time). Figure 4 shows the total hourly deletions for different kinds of posts (on a log scale) from 20 July to 8 September 2012. Both "general deletions" and "system deletions" happen even very late at night.

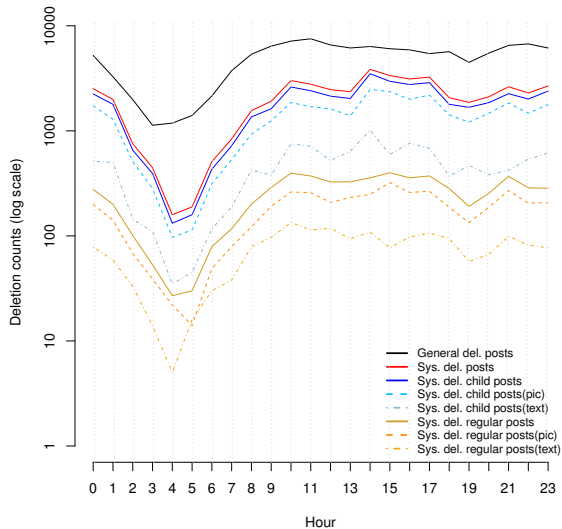

Figure 4: **Post deletion amounts over 24 hours.**

So do the censors respond as quickly during the night as during day hours? We plotted the median lifetime of the posts as a function of their deletion time in Figure 5. The morning-hour spike suggests that the censors are behind in the morning, both catching up on overnight posts and dealing with a fresh influx of posts from morning risers. They catch up by late morning or early afternoon.

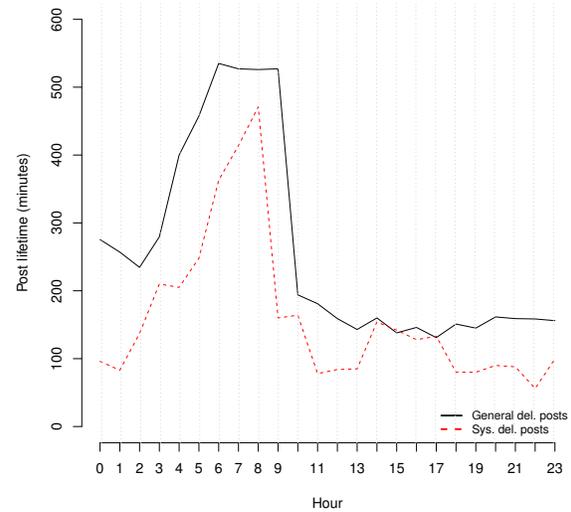

Figure 5: **Post lifetime *vs.* deletion time of the day.**

From Figure 4 and Figure 5 it is clear that, while a significant fraction of the censors seem to work during regular work hours, many do not.

## 6.3 Synthesis

Based on everything we have seen and observed, we can begin to understand how Weibo censorship works. Clearly, they are using a strong degree of automation to help them delete posts that have been declared sensitive. It is also clear that this process is relatively "loose," in the sense that there are few sharp rules that define what gets deleted *vs.* what is allowed to remain. Given the long-tailed distribution that we observe in post lifetimes prior to censorship, it is clear that some posts are not considered a high priority for censorship, such as if two friends start conversing with each other using a new neologism, euphemism, or other coinage that would otherwise be censorship-worthy. However, when those new terms spread and grow, they are censored both proactively and retroactively.

This suggests that Weibo is trying to strike a balance between satisfying the legal requirements within which it operates and the costs of running a fine-grained instrument of political censorship. Weibo must conduct *just enough* censorship to satisfy government regulations without being so intrusive as to discourage users from using their service. Among other issues, they must surely be deeply concerned with false positives. If truly innocu-ous posts disappeared with any regularity, Weibo's users might defect to a competing service.

It is unclear the extent to which Weibo is using natural language processing (NLP) algorithms to aid in their work, versus having a stable of censors watching for things to go viral and then using search tools to stamp them out. Certainly, our use of fairly simple NLP techniques helped reduce the workload of analyzing trending topics, so comparable techniques may well be in use by Weibo. NLP techniques in a censor's hands can be thought of as a "force multiplier," but it is unclear whether they fundamentally change the game. Consider, with English-language spam emails, the degree to which spammers will try to evade automated spam classification systems. These techniques and more could well be applied to automated or manual rewriting of postings, with the intent of avoiding automated censorship. The results might not be as easy to read, but humans will likely have an advantage at reading jumbled text, at least until NLP algorithms are extended to deal with them. Conversely, NLP techniques can cluster together related terms, assisting censors to overcome such techniques. At least so far, we have not seen evidence of any sort of arms race between increasingly sophisticated ways to avoid censorship and increasingly powerful censorship techniques.

In many ways, Internet censorship is related to intrusion detection. When our results in this paper are compared to related work (see Section 1), including both IP-layer filtering and application-level censorship, a picture of Internet censorship in China emerges where "defense-in-depth" is taken to a new level. Intrusion detection research has long focused on issues such as false positive *vs.* false negative tradeoffs, viral spreading patterns, polymorphic content, and the distinction between different layers of abstraction (such as IP packets *vs.* application-layer data). The so-called "Great Firewall of China" and the accompanying application-layer censorship that China's domestic web services, such as Weibo, carry out afford us an opportunity to study a real, national scale intrusion detection system.

### 6.4 Major caveats

The most important caveat to keep in mind when interpreting our results is that we collected posts from a very specific core set of users, built up from a "seed" group of users who post about sensitive topics, which we call the "user timeline." Unless otherwise noted, such as when results are from the public timeline, all results in this paper are from the user timeline and therefore might be biased by the differences between this core set of users and the average Weibo user. All deletion rates, deletion times, *etc.* must be interpreted in this light. In other words, our sample users should not be considered to be representative of the general population of Weibo.

Another important caveat is that our system does not detect post deletions in the user timeline if the post deleted is not one of the 50 most recent posts by the user (see Section 3). This may affect our results about the distribution of post deletions over time in Section 4.

## 7 Conclusion

Our research found that deletions happen most heavily in the first hour after a post has been made (see Figure 1). Especially for original posts that are not reposts, most deletions occur within 30 minutes, accounting for 30% of the total deletions of such posts. Nearly 90% of the deletions of such posts happen within the first 24 hours of the post.

With respect to the hypotheses enumerated in Section 4, we make the following conclusions:

- Hypothesis 1: The Weibo system keeps more than one keyword list, where each list triggers a different kind of censorship behavior.

- Hypothesis 2: The clear downward trend in Figure 2 could be evidence that certain users are flagged for closer scrutiny, but we have not ruled out other causes in this paper.

- Hypothesis 3: Figure 3 shows that over 82% of reposted posts have a standard deviation of less than 5 minutes deletion time, meaning that a sensitive post is detected and then most of the other posts in a chain of reposts are then deleted.

- Hypothesis 4: As described in Section 4, using the methods described in Section 5 we find that topics that were trends in the user timeline and were also, according to the public timeline, hot topics in public discussion as a whole about events that happened during our month of data collection (Qidong, Qian Yunhui, Beijing Rainstorms, Diaoyu islands, and a group sex scandal) had very short lifetimes. Recall that the deleted posts in the user timeline included themes related to state power (`Beijing, government, China, country, policeman, and people`). This suggests that such broadly discussed topics are targeted with more censorship resources to limit certain kinds of discussion about the events.

Future work may reveal many mechanisms beyond those we described here, and many different strategies that Weibo uses to prioritize what content to delete. Our results suggest that Weibo employs a distributed, heterogeneous strategy for censorship that has a great amount of "defense-in-depth."

One aspect of censorship that is not considered in our analysis, but would be an interesting topic for future work, is the interactions between social media and traditional media. Leskovec *et al.* [17] gives an interesting analysis of the interplay between blogs and traditional media during the 2008 U.S. Presidential election. Traditional media relevant to Weibo may include the state-run media that is heavily censored, or off-shore news outlets that are uncensored but limited in availability and sometimes offset from China's news cycles by timezone differences.

## 8 Acknowledgments

We would like to thank the anonymous reviewers and our shepherd, Nikita Borisov, for helpful feedback. We owe our deepest gratitude to Professors Stephanie Forrest, Christopher Bronk, and George Luger for their feedback and comments, and for encouraging us to go forward. We are also grateful to Ben Edwards for insightful discussions about potential future work. This material is based upon work supported by the National Science Foundation under Grant Nos. #0844880, #0905177, #1017602. Jed Crandall is also supported by the Defense Advanced Research Projects Agency CRASH program under grant #P-1070-113237.